\documentclass[RNAAS]{aastex62}

\usepackage{graphicx}	
\usepackage{amsmath}	

\usepackage{txfonts}
\usepackage[T1]{fontenc}
\usepackage{ae,aecompl}
\graphicspath{{./}{figures/}}

\begin{document}

\title{An Argument for a Kilometer-Scale Nucleus of C/2019 Q4}

\email{amir.siraj@cfa.harvard.edu, aloeb@cfa.harvard.edu}

\author{Amir Siraj}
\affil{Department of Astronomy, Harvard University, 60 Garden Street, Cambridge, MA 02138, USA}

\author{Abraham Loeb}
\affiliation{Department of Astronomy, Harvard University, 60 Garden Street, Cambridge, MA 02138, USA}

\begin{abstract}

C/2019 Q4 (Borisov), discovered by Gennady Borisov on August 30, 2019, is tentatively an interstellar comet. We show that a kilometer-scale nucleus would provide consistency with both ‘Oumuamua and CNEOS-2014-01-08, resulting in a single power-law distribution with an equal amount of mass per logarithmic bin of interstellar objects.

\end{abstract}

\section{Introduction}
`Oumuamua was the first interstellar object detected in the Solar System by Pan-STAARS \citep{Meech2017, Micheli2018}. Its size was estimated to be 20m - 200m, based on Spitzer Space Telescope constraints on its infrared emission given its temperature \citep{Trilling2018}. CNEOS 2014-01-08 is tentatively the first interstellar meteor discovered larger than dust, with $\sim 1\mathrm{\; m}$ in size \citep{Siraj2019}\footnote{Still in ApJL peer-review.}.

C/2019 Q4 (Borisov), discovered by Gennady Borisov on August 30, 2019, at the Crimean Astrophysical Observatory, is tentatively an interstellar comet \citep{Guzik2019}. According to a JPL press release, the radius of the nucleus has been estimated by Karen Meech and her team at the University of Hawaii to be $1 - 8 \mathrm{\; km}$.\footnote{https://www.jpl.nasa.gov/news/news.php?feature=7498} In this \textit{Note}, we show that a kilometer-scale nucleus would provide consistency with both `Oumuamua and CNEOS-2014-01-08, resulting in a single power-law distribution with an equal amount of mass per logarithmic bin of interstellar objects.

\section{Orbit}
As C/2019 Q4 has yet to cross the plane of the ecliptic, previous planetary encounters in the Solar System are impossible. In addition, it is unlikely to have received a substantial non-gravitational kick from cometary exhaust due to its large distance from the Sun as well as its substantial size.

The orbit of C/2019 Q4 implies an excess heliocentric velocity of $\sim 30 \; \mathrm{km \; s^{-1}}$ \citep{Guzik2019}. The heliocentric incoming velocity at infinity of the meteor in right-handed Galactic coordinates is $v_{\infty}\mathrm{(U, V, W) \approx (21, -23, 1)\;}$ $\mathrm{km\;s^{-1}}$, which is $\sim 35\; \mathrm{km\;s^{-1}}$ away from the velocity of the Local Standard of Rest (LSR), $\mathrm{(U, V, W)_{LSR}}$ $= (-11.1, -12.2, -7.3)\;$ $\mathrm{km\;s^{-1}}$ \citep{Schonrich2010}. The object's speed likely reflects the speed of the parent star if it originated from the star's Oort cloud.

Its orbit is inclined $\sim 15^{\circ}$ to the Galactic plane, suggesting a possible origin in the Galactic disk. If C/2019 Q4 comes from the Galactic disk, then a stellar origin from a distance $\gtrsim 1 \; \mathrm{kpc}$ is most likely. Given the large distance to the parent star relative to the Galactic scale height, it is unlikely to find the parent star through backward trajectory extrapolation.

\section{Implied Ejected Mass}

A rough, order-of-magnitude estimate of number density assumes a search time of a few years with detection coverage out to $\sim 3 \mathrm{\; AU}$ (giving a detection volume of $\sim 10^2 \mathrm{\; AU^3}$), which an implied number density of a few $10^{-3{^{+0.75}_{-1.5}}} \mathrm{\; AU^{-3}}$ or a few $10^{13{^{+0.75}_{-1.5}}} \mathrm{\; pc^{-3}}$ (given 95\% Poisson uncertanties). Such a number density is a factor of $\sim 10^{-2}$ of that implied by `Oumuamua \citep{Do2018}, suggesting a power law index similar to Solar System bodies at those size scales \citep{Ivezic2001}.

\begin{figure}
  \centering
  \includegraphics[width=.7\linewidth]{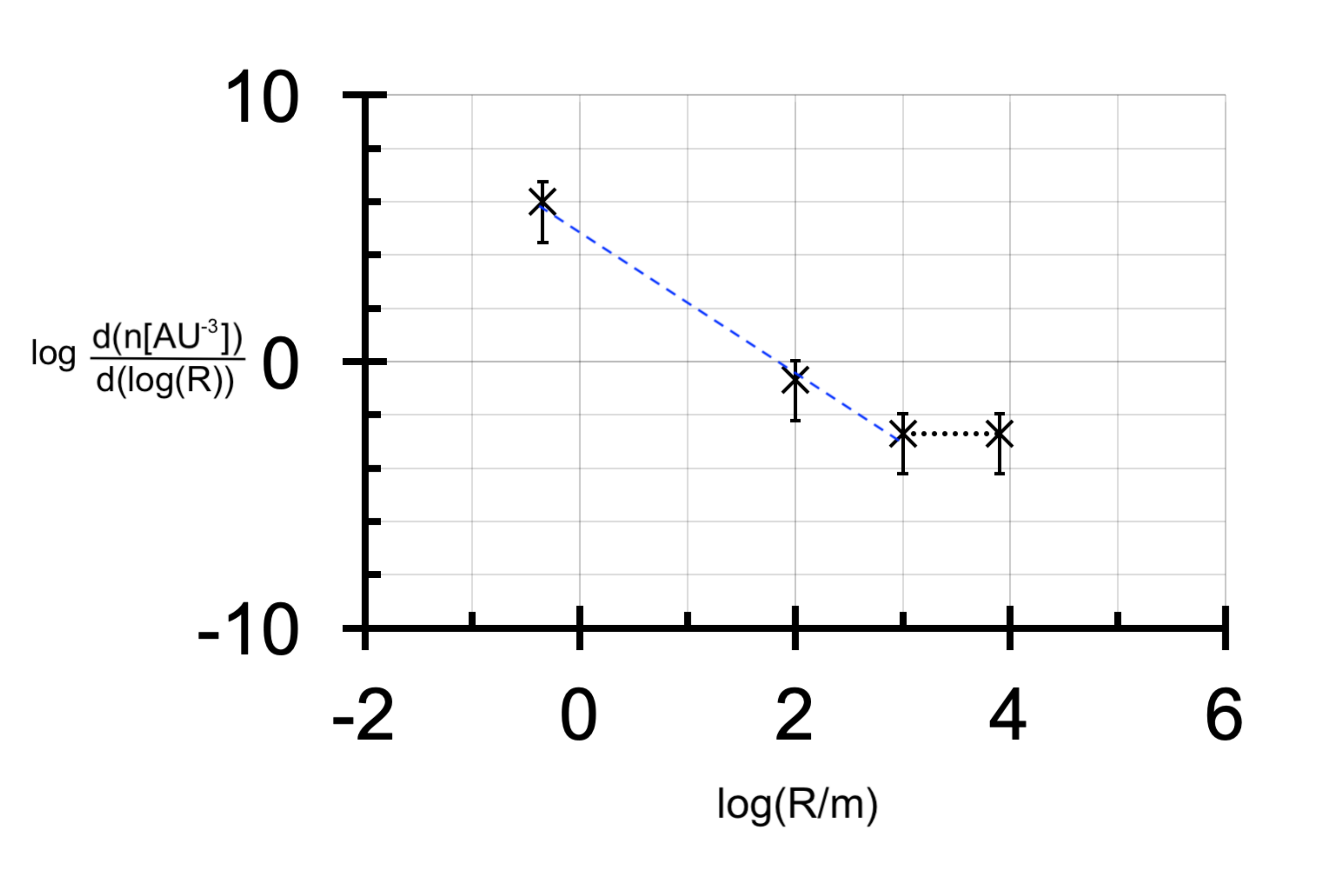}
    \caption{Size distribution of interstellar objects, given `Oumuamua, CNEOS 2014-01-08, and C/2019 Q4, each with 95\% Poisson confidence intervals. The two rightmost points indicate C/2019 Q4 with 1 km and 8 km radii, respectively. The blue dotted line indicates a slope of $\sim-3$, implying an equal amount of mass per logarithmic bin of interstellar objects.}
    \label{fig:3}
\end{figure}

Assuming a typical cometary density of $0.6 \mathrm{\; g \; cm^{-3}}$ and the largest plausible radius of $8 \mathrm{\; km}$, C/2019 Q4 would have a mass of $10^{18} \mathrm{\; g}$. Taken together with our estimate of implied number density, this implies a mass density of $\sim 10^{31{^{+0.75}_{-1.5}}} \mathrm{\; g \; pc^{-3}}$, or $\sim10^{1{^{+0.75}_{-1.5}}}$ Jupiter masses ejected per star. This density is $\sim 10^{3{^{+0.75}_{-1.5}}}$ higher than that expected of Oort clouds \citep{Weissman1983}. This would constitute an extremely surprising implication for the amount of ejected mass in the form of large comets. If the size of the nucleus were of order $\sim 1 \mathrm{\; km}$, the required ejected mass per star would fall to the expected range for Oort clouds (a few $10^{28} \mathrm{\; g}$), comparable to the ejected mass implied by both `Oumuamua and CNEOS-2014-01-08 for their respective logarithmic bins \citep{Siraj2019}. Figure~\ref{fig:3} shows the size distribution of interstellar objects with `Oumuamua, CNEOS 2014-01-08, and C/2019 Q4 (for radii of 1 km and of 8 km), with appropriate Poisson uncertainties.

Further measurements will be crucial for constraining the scale of the nucleus of C/2019 Q4 and the corresponding implications for ejected mass. We recommend spectroscopy of the cometary gases throughout the monitoring of C/2019 Q4 in the future. A combination of precision astrometry and spectroscopy of C/2019 Q4 will reveal its true origin, as well as its composition and internal structure, all of which are crucial for understanding the formation and constituents of exo-planetary systems.

\section*{Acknowledgements}
This work was supported in part by a grant from the Breakthrough Prize Foundation. 

\end{document}